\documentclass{pasj01}
\usepackage{lscape}
\usepackage{comment}
\usepackage{ccaption}
\usepackage {threeparttable}
\usepackage{color}
\draft 
\Received{$\langle$reception date$\rangle$}
\Accepted{$\langle$acception date$\rangle$}
\Published{$\langle$publication date$\rangle$}

\begin{document}

\title{Massive star formation in the Carina nebula complex and Gum 31 -- II. a cloud--cloud collision in Gum 31}
\author{Shinji FUJITA$^{1*}$}
\author{Hidetoshi SANO$^{2}$}%
\author{Rei ENOKIYA$^{3}$}%
\author{Katsuhiro HAYASHI$^{4}$}%
\author{Mikito KOHNO$^{5}$}%
\author{Kisetsu TSUGE$^{6}$}%
\author{Kengo TACHIHARA$^{7}$}%
\author{Atsushi Nishimura$^{8}$}%
\author{Akio OHAMA$^{7}$}%
\author{Yumiko YAMANE$^{7}$}%
\author{Takahiro OHNO$^{7}$}%
\author{Rin I. YAMADA$^{7}$}%
\author{Yasuo FUKUI$^{7}$}%

\altaffiltext{1}{Department of Physical Science, Graduate School of Science, Osaka Prefecture University 1-1 Gakuen-cho, Naka-ku, Sakai, Osaka 599-8531, Japan}
\altaffiltext{2}{National Astronomical Observatory of Japan, Mitaka, Tokyo, 181-8588, Japan}
\altaffiltext{3}{Department of Physics, Institute of Science and Technology, Keio University, 3-14-1 Hiyoshi, Kohoku-ku, Yokohama, Kanagawa 223-8522, Japan}
\altaffiltext{4}{Institute of Space and Astronautical Science (ISAS) Japan Aerospace Exploration Agency (JAXA), 3-1-1 Yoshinodai, Chuo-ku, Sagamihara, Kanagawa 252-5210, Japan}
\altaffiltext{5}{Astronomy Section, Nagoya City Science Museum, 2-17-1 Sakae, Naka-ku, Nagoya, Aichi 460-0008, Japan}
\altaffiltext{6}{Dr. Karl Remeis Observatory, Erlangen Centre for Astroparticle Physics, Friedrich-Alexander-Universit\"{a}t Erlangen-N\"{u}rnberg, Sternwartstra$\beta$e 7, 96049 Bamberg, Germany}
\altaffiltext{7}{Department of Astrophysics, Nagoya University, Furo-cho, Chikusa-ku, Nagoya, Aichi, 464-8602, Japan}
\altaffiltext{8}{Institute of Astronomy, the University of Tokyo, 2-21-1 Osawa, Mitaka, Tokyo 181-0015, Japan}

\email{fujita@p.s.osakafu-u.ac.jp}


\maketitle

\begin{abstract}
We present the results of analyses of the $^{12}$CO ($J$=1--0), $^{13}$CO ($J$=1--0), and $^{12}$CO ($J$=2--1) emission data toward Gum 31.
Three molecular clouds separated in velocity were detected at $-25$, $-20$, and $-10$ \,km\,s$^{-1}$. 
The velocity structure of the molecular clouds in Gum 31 cannot be interpreted as expanding motion.
Two of them, the $-25$\,km\,s$^{-1}$ cloud and the $-20$\,km\,s$^{-1}$ cloud, are likely associated with Gum 31, because their $^{12}$CO ($J$=2--1)/$^{12}$CO ($J$=1--0) intensity ratios are high.
We found that these two clouds show the observational signatures of cloud--cloud collisions (CCCs): a complementary spatial distribution and a V-shaped structure (bridge features) in the position--velocity diagram. 
In addition, their morphology and velocity structures are very similar to the numerical simulations conducted by the previous studies.
We propose a scenario that the $-25$\,km\,s$^{-1}$ cloud and the $-20$\,km\,s$^{-1}$ cloud were collided and triggered the formation of the massive star system HD 92206 in Gum 31. 
This scenario can explain the offset of the stars from the center and the morphology of Gum 31 simultaneously.
The timescale of the collision was estimated to be $\sim$\,1\,Myr by using the ratio between the path length of the collision and the assumed velocity separation.
This is consistent with that of the CCCs in Carina Nebula Complex in our previous study. 
\end{abstract}

\section{Introduction}
Many recent observational studies of star forming region have revealed the triggering of massive star formations via cloud--cloud collisions (CCCs). 
The two observational signatures---the complementary spatial distributions and bridging features between the colliding clouds---have been suggested as evidence (e.g., \cite{2020arXiv200905077F}).
Such evidence was detected in CCC regions both in the Galactic (e.g., \cite{2009ApJ...696L.115F, 2010ApJ...709..975O, 2014ApJ...780...36F, 2015ApJ...806....7T, 2016ApJ...820...26F, 2017arXiv170605664F, 2017PASJ...69L...5F, 2017ApJ...835..142T, 2017ApJ...840..111T, 2018PASJ...70S..49E, 2018PASJ...70S..48H, 2018PASJ...70S..50K, 2018PASJ...70S..42N, 2018PASJ...70S..43S, 2019PASJ..tmp...46F, 2019ApJ...872...49F, 2019ApJ...886...14F, 2018PASJ..tmp..121T, 2018PASJ...70S..51T, 2020arXiv200905077F, 2020PASJ..tmp..238F}) and extragalactic (e.g., \cite{2019ApJ...886...15T, 2020PASJ..tmp..209S}) objects.

In \citet{2020PASJ..tmp..238F} (hereafter Paper I), we analysed a CO dataset towards the Carina Nebula Complex (CNC). 
Four discrete molecular clouds at velocities $-27$, $-20$, $-14$, and $-8$\,km\,s$^{-1}$ were found. 
Most of them might be associated with the star clusters in the CNC, because their $^{12}$CO ($J$=2--1)/$^{12}$CO ($J$=1--0) intensity ratios are high and their distributions correspond to the {\it Spitzer} 8-$\mu$m distributions.
In addition to this, we found that these clouds show the observational signatures of a CCC.
Furthermore, we also found that the SiO emission (a tracer of a shocked molecular gas) is significantly enhanced between the colliding clouds. 
For these results, we propose a scenario that the formation of massive stars in the clusters in the CNC was triggered by CCCs with a timescale of $\sim$\,1\,Myr.

\begin{figure*}[htbp]
  \begin{center}
  \includegraphics[width=16cm]{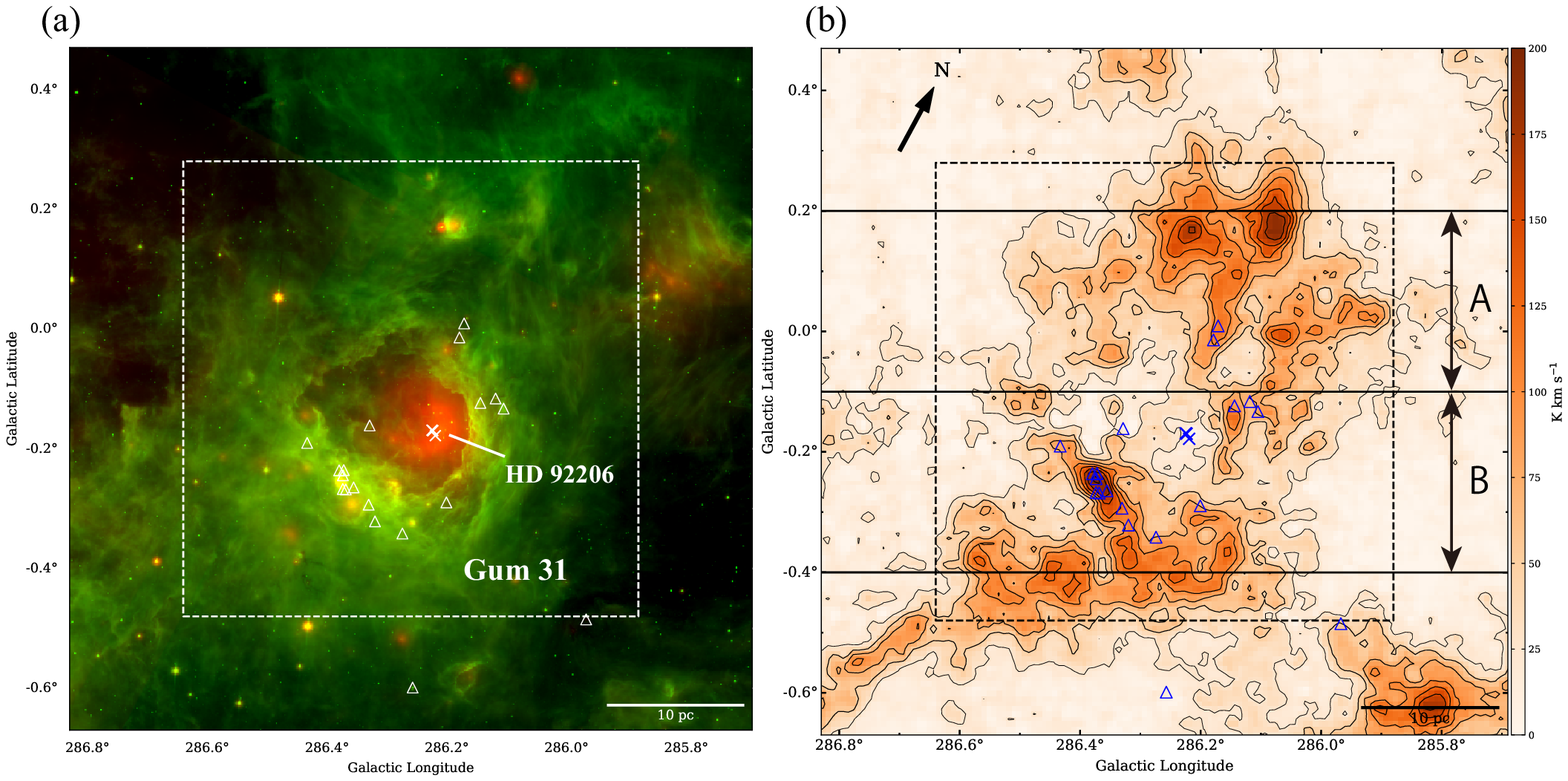}
  \end{center}
  \caption{(a) Composite image of the WISE 22\,$\mu$m (red) and {\it Spitzer}/GLIMPSE 8\,$\mu$m (green) towards Gum 31.  (b) The $^{12}$CO ($J$=1--0) integrated-intensity map taken from the Mopra archive (\cite{Reb16, Reb17}). Integrated velocity range is from -35 to 10\,km\,s$^{-1}$. The black contours are plotted 25 ($\sim 10\sigma$), 50, 75, ..., 200\,K\,km\,s$^{-1}$. The black dotted rectangle indicates the area observed with NANTEN2. The crosses  represent the multiple O-stars HD 92206 (\cite{1988A&AS...76..427M}). The triangles represent young stellar objects (\cite{2013A&A...552A..14O}). }\label{fig:RGB}
\end{figure*}

In this paper, we focus on Gum 31 (E127 in \citet{2019PASJ...71....6H}).
Gum 31 located $\sim 50$\,pc from the CNC on the sky plane.
The distance to Gum 31 has been discussed and determined to be 2.5\,$\pm$0.3\,kpc (\cite{2010MNRAS.402...73B}). 
We adopt the value of 2.5\,kpc in this paper.
Figure\,\ref{fig:RGB}(a) shows a composite color image of the WISE 22-$\mu$m (red) and {\it Spitzer}/GLIMPSE 8-$\mu$m (green) emissions around Gum 31. 
It is ionized by the young stellar cluster NGC 3324 (e.g., \cite{1982AJ.....87.1300W}) including the double or multiple O-type stars in HD 92206 \textcolor{black}{[HD 92206A at $(l,\,b)=(286\fdg223,\,-0\fdg170)$, HD 92206B at $(l,\,b)=(286\fdg225,\,-0\fdg169)$, and HD 92206C (CD-57$\fdg$3378) at $(l,\,b)=(286\fdg219,\,-0\fdg178)$] (\cite{1988A&AS...76..427M})}. 
Gum 31 is relatively a small and resembles to the {\it Spitzer} bubbles (\cite{2006ApJ...649..759C, 2007ApJ...670..428C}), although the radius is somewhat larger ($\sim$10\,pc) than is typical for those bubbles. 
\textcolor{black}{Gum31 is connected to the central parts of the CNC in the Herschel far-infrared maps (\cite{2013A&A...552A..14O}) and CO maps (\cite{2005ApJ...634..476Y, Reb16}).
In addition, the distance to the CNC is estimated to be $\sim$2.3\,kpc (e.g., \cite{All93, Smi06a}) and thus Gum31 is probably associated with the CNC.}

\citet{2008A&A...477..173C} investigated the ionized, neutral, and CO gas in the environs of Gum 31 and found that the H{\sc i} structure is 10.0 $\pm$ 1.7\,pc in radius, has a neutral mass of 1500 $\pm$ 500 $M_{\odot}$, and is expanding at 11\,km\,s$^{-1}$.
They concluded that the expansion of the H{\sc ii} region triggered stellar formation in the molecular shell, although the angular resolution of their CO data is insufficient ($\sim \,2'.7$).
\textcolor{black}{Detailed gas observations and analyses in such a young [\textcolor{black}{the age of the cluster NGC 3324 is estimated to be } $< \, 3$\,Myr (\cite{2001A&A...371..107C})] star forming region are important in revealing the mechanism of massive star formation. }

\section{Dataset}

\subsection{Mopra}\label{sec:dat_m}

As in Paper I, we used the $^{12}$CO and $^{13}$CO ($J$=1--0) archived datasets obtained with the Mopra 22-m telescope (\cite{Bur13, Bra15, Reb16, Reb17}), the beam size and velocity resolution of which are $\sim$35$''$ and $\sim$0.09\,km\,s$^{-1}$, respectively.
The rms noise levels for the $^{12}$CO and $^{13}$CO ($J$=1--0) datasets are $\sim$4.0\,K and $\sim$1.9\,K per channel on the $T_{\rm mb}$ scale, respectively, at a velocity grid of 0.09\,km\,s$^{-1}$. 

\subsection{NANTEN2}\label{sec:dat_n}

The $^{12}$CO ($J$=2--1) data toward Gum 31 were obtained with the NANTEN2 4-m millimeter/sub-millimeter telescope at Atacama, Chile in 2015 October. 
The beam size of NANTEN2 at $\sim$230\,GHz corresponds to $\sim$90$''$, and we adopted the on-the-fly (OTF) mapping mode with Nyquist sampling. 
The backend and frontend system and calibration are the same as in Paper I.
A typical uncertainty in the intensity is $\sim$20\%, and the typical rms noise level for the $^{12}$CO ($J$=2--1) data is $\sim$0.3\,K per channel on the $T_{\rm mb}$ scale at the velocity grid of 0.5\,km\,s$^{-1}$.

\section{Results}\label{sec:res}

Figure\,\ref{fig:RGB}(b) shows the $^{12}$CO ($J$=1--0) integrated-intensity distributions (Mopra) with the velocity range from $-35$ to $10$\,km\,s$^{-1}$.
It has a roughly ring-shaped structure, and the double or multiple O-type stars in HD 92206 has been identified (e.g., \cite{1988A&AS...76..427M, 2004ApJS..151..103M}) at a location that is offset eastward from the center.
We estimated that the maximum column density of the molecular gas is $1\,\times \,10^{23}$\,cm$^{-2}$ and the total molecular gas mass \textcolor{black}{(within the dotted square in Figure\,\ref{fig:RGB})} is about $1.4\,\times \,10^{5}$\,$M_{\odot}$ by using the same method as in Paper I.

\begin{figure*}[htbp]
  \begin{center}
  \includegraphics[width=16cm]{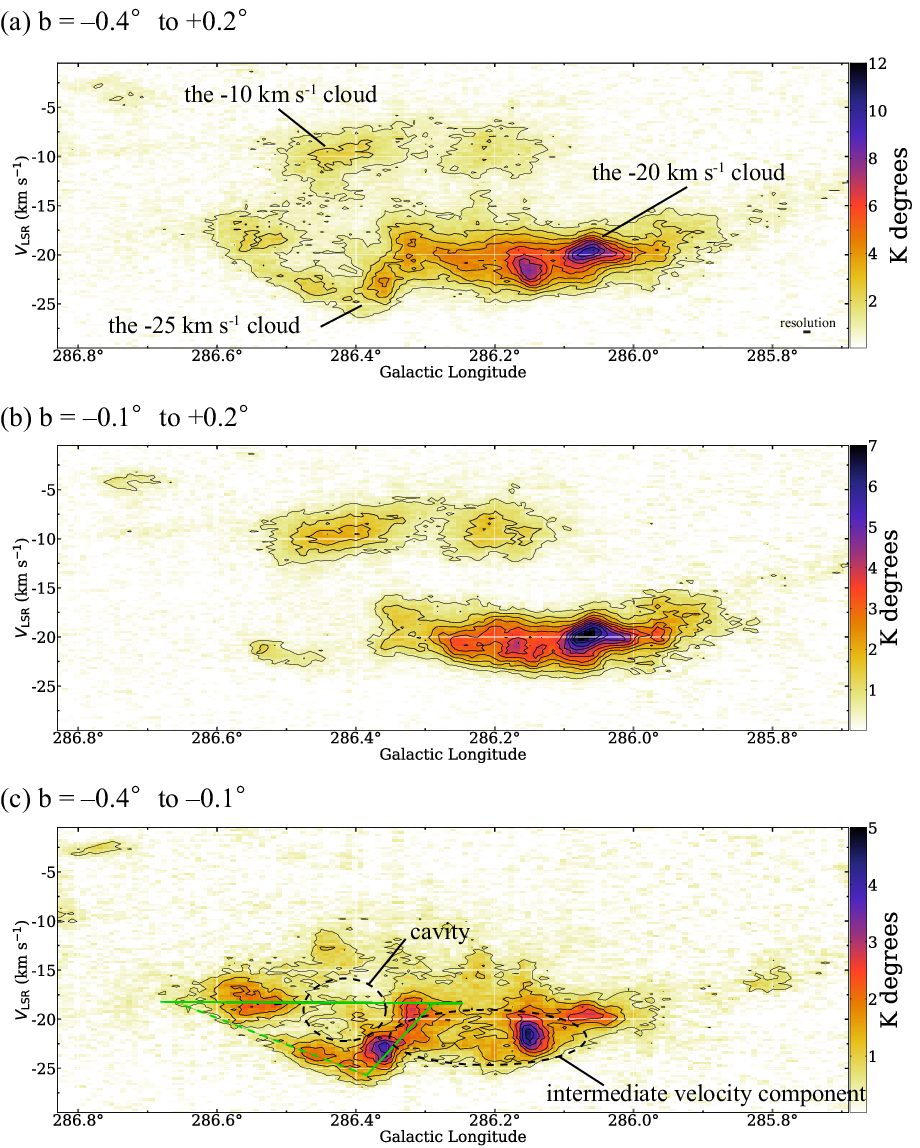}
  \end{center}
  \caption{(a) Galactic longitude -- Velocity ($l$--$v$) diagram of the $^{12}$CO ($J$=1--0) emissions integrated from -0\fdg4 to +0\fdg2. (b) The same as (a), but integrated from -0\fdg1 to +0\fdg2 (A in Figure\,\ref{fig:RGB}(b)). (c) The same as (a), but integrated from -0\fdg4 to -0\fdg1. (B in Figure\,\ref{fig:RGB}(b))}\label{fig:lvs}
\end{figure*}

Figure\,\ref{fig:lvs} shows the $l$--$v$ diagrams of the $^{12}$CO ($J$=1--0) toward Gum 31 with the integrated ranges (a) from $-0$\fdg$4$ to $+0$\fdg$2$, (b) from $-0$\fdg$1$ to $+0$\fdg$2$, and (c) from $-0$\fdg$4$ to $-0$\fdg$1$.
In Figure\,\ref{fig:lvs}(b), in the northern part, we detected two clouds separated in velocity and centered at $\sim\,$$-20$ and $\sim\,$$-10$\,km\,s$^{-1}$ (hereinafter called the $-20$\,km\,s$^{-1}$ cloud and the $-10$\,km\,s$^{-1}$ cloud, respectively).
On the other hand, in Figure\,\ref{fig:lvs}(c), there are two velocity components separated in velocity and centered at $\sim\,$$-25$ (hereinafter called the $-25$\,km\,s$^{-1}$ cloud) and $\sim\,$$-20$\,km\,s$^{-1}$ with a velocity gradient. 
In addition, there is also an intermediate velocity component in between.

\begin{figure*}[htbp]
  \begin{center}
  \includegraphics[width=16cm]{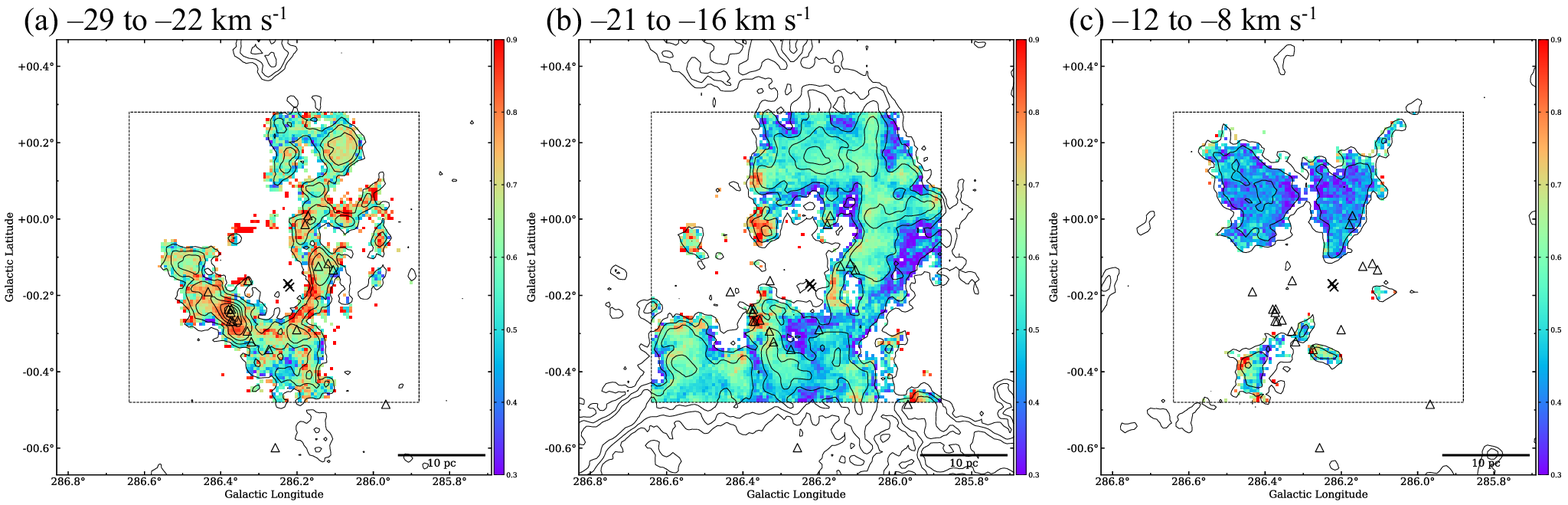}
  \end{center}
  \caption{The integrated-intensity ratio $^{12}$CO ($J$=2--1)/$^{12}$CO ($J$=1--0) ($R^{12}_{2-1/1-0}$) map for the velocity range (a) from $-29$\,km\,s$^{-1}$ to $-22$\,km\,s$^{-1}$, (b) from $-21$\,km\,s$^{-1}$ to $-16$\,km\,s$^{-1}$, and (c) from $-12$\,km\,s$^{-1}$ to $-8$\,km\,s$^{-1}$, respectively. The symbols are the same as in Figure\,\ref{fig:RGB}. }\label{fig:integs}
\end{figure*}

Figure\,\ref{fig:integs} shows the integrated-intensity ratios $^{12}$CO ($J$=2--1)/$^{12}$CO ($J$=1--0) (hereinafter denoted by $R^{12}_{2-1/1-0}$) (color image) and the integrated-intensity distributions of the $^{12}$CO ($J$=1--0) emission (contours) for the three clouds. 
The $R^{12}_{2-1/1-0}$ of the $-25$\,km\,s$^{-1}$ cloud and the $-20$\,km\,s$^{-1}$ cloud are high ($>$0.8) at the edge of the clouds, indicating that they are heated by the surrounding H{\sc ii} region in the same way as the cloud in the CNC.
In contrast, the $R^{12}_{2-1/1-0}$ of the $-10$\,km\,s$^{-1}$ cloud is low ($\sim \,0.4$).
For these reasons, it is probable that the former two are associated with the H{\sc ii} region, and the latter is not.

\section{Discussion}\label{sec:Dis}
The observations with Mopra and NANTEN2 have shown that the two CO clouds, the $-25$\,km\,s$^{-1}$ cloud and the $-20$\,km\,s$^{-1}$ cloud, are both associated with Gum 31.
The molecular mass of the two are $4.9\,\times \,10^4$\,$M_{\odot}$ (from $-27$\,km\,s$^{-1}$ to $-21$\,km\,s$^{-1}$) and $2.3\,\times \,10^4$\,$M_{\odot}$ (from $-21$\,km\,s$^{-1}$ to $-15$\,km\,s$^{-1}$), respectively, within a radius of 15\,pc centered on Gum 31.
By tentatively assuming the most plausible viewing angle of the relative motion of the clouds to the line-of-sight is 60$^{\circ}$ (\cite{2019PASJ..tmp...46F}), the relative velocity between the two is 10\,km\,s$^{-1}$, which requires the total mass of $\frac{rv^2}{2G}\,=\,1.7\,\times \,10^5$\,$M_{\odot}$ to bind the two clouds gravitationally within 15\,pc (where $r$ and $G$ are the distance between the two clouds and the gravitational constant, respectively).
This is a factor of 2 larger than the total molecular mass of the two clouds derived, indicating that the coexistence of the two clouds cannot be understood as a gravitationally bound system.
\textcolor{black}{Therefore, the velocity structure of the molecular gas in Gum 31 is thought to be in the middle of or as a result of some dynamic event. }

\subsection{Expanding shell model}
\begin{figure}[htbp]
  \begin{center}
  \includegraphics[width=16cm]{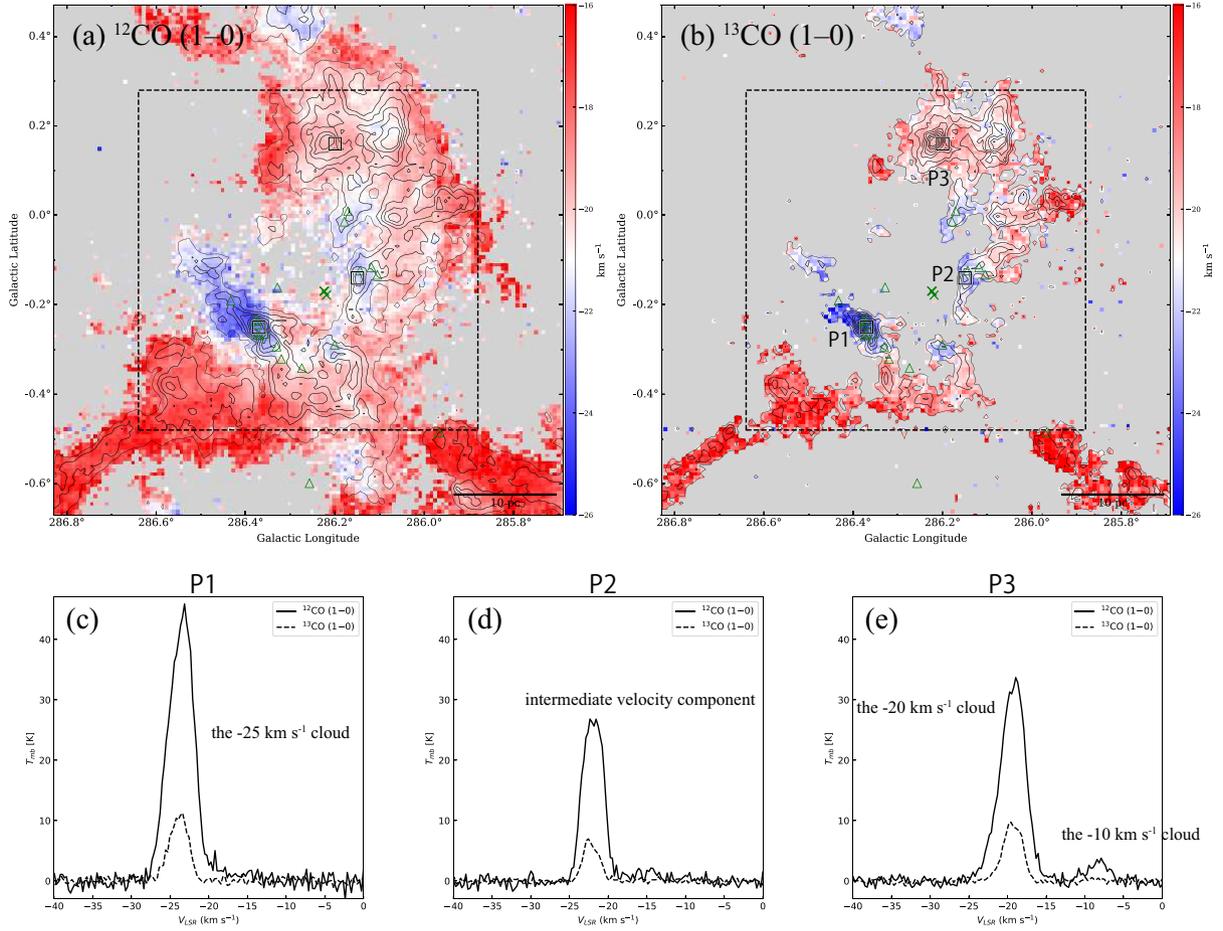}
  \end{center}
  \caption{(a) Map of the 1st moment of the $^{12}$CO ($J$=1--0) emission between $-27$\,km\,s$^{-1}$ and $-15$\,km\,s$^{-1}$. The contours show the moment 0 (integrated-intensity) of the $^{12}$CO ($J$=1--0) emission. The contours are plotted every 25\,K\,km\,s$^{-1}$ from 25\,K\,km\,s$^{-1}$ ($\sim 10\sigma$). The symbols are the same as in Figure\,\ref{fig:RGB}. \textcolor{black}{(b) The same as (a), but for the $^{13}$CO ($J$=1--0). (c--e) The spectra of the $^{12}$CO ($J$=1--0) and the $^{13}$CO ($J$=1--0) at the position P1, P2, and P3, respectively, shown by square makers. }}\label{fig:mom1}
\end{figure}

A possible explanation of the velocity components and the morphology of Gum 31 is the feedback effects from the exciting stars in HD 92206, such as the stellar wind, UV radiation, and pressure of expanding H{\sc ii} region.
However, HD 92206 stars are offset from the center of the bubble-like structure of Gum 31.
In addition, in Figure\,\ref{fig:lvs}(a) we can not see clear expanding motion, namely an ellipse in the $p$--$v$ diagram, of the cloud either in the northern or the southern part of Gum 31.
Figures\,\ref{fig:mom1}(a)(b) shows the first moment of the $^{12}$CO ($J$=1--0) emissions \textcolor{black}{and the $^{13}$CO ($J$=1--0) emissions, respectively,} with a velocity range from $-27$\,km\,s$^{-1}$ to $-15$\,km\,s$^{-1}$. 
It is clear that the insider cloud is blueshifted (the $-25$\,km\,s$^{-1}$ cloud) and the outsider is redshifted (the $-20$\,km\,s$^{-1}$ cloud).
If the cloud is isotropically expanding, such a velocity gradient should not be seen.
For these reasons, molecular gas in Gum 31 is likely to not have an expanding structure.

\subsection{An Alternative: Cloud--cloud collision}

Paper I suggests complex CCCs in the CNC.
\textcolor{black}{The existence of different velocity components, relatively weak CO emissions at intermediate velocities between two clouds (a bridge feature) in a $p$--$v$ diagram, and a complementary spatial distribution were reported as evidence of CCCs.
The combination of the two characteristics of a bridge feature and a complementary spatial distribution made the $p$--$v$ diagram V-shaped.}

\begin{figure}[htbp]
  \begin{center}
  \includegraphics[width=8cm]{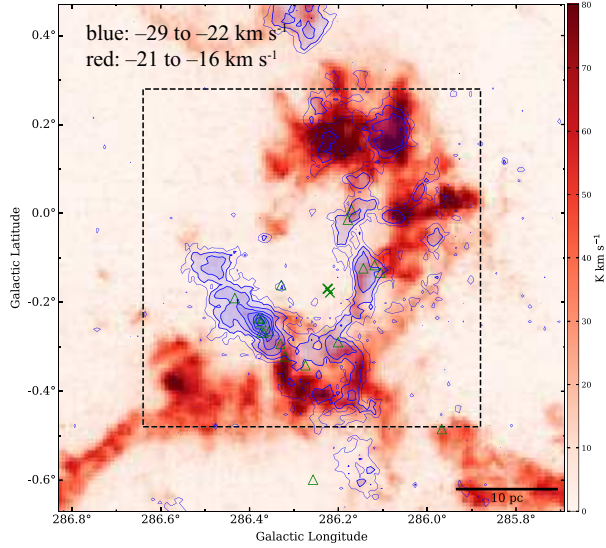}
  \end{center}
  \caption{The red and blue show the integrated-intensity distributions of the $^{12}$CO ($J$=1--0) emissions toward Gum 31. The each integrated velocity ranges are shown in the top left corner of the figure. The contour levels are 8, 16, 32, 64, 96, 128, and 160\,K\,km\,s$^{-1}$. The symbols are the same as in Figure\,\ref{fig:RGB}. }\label{fig:red_blue}
\end{figure}

\begin{figure*}[htbp]
  \begin{center}
  \includegraphics[width=14cm]{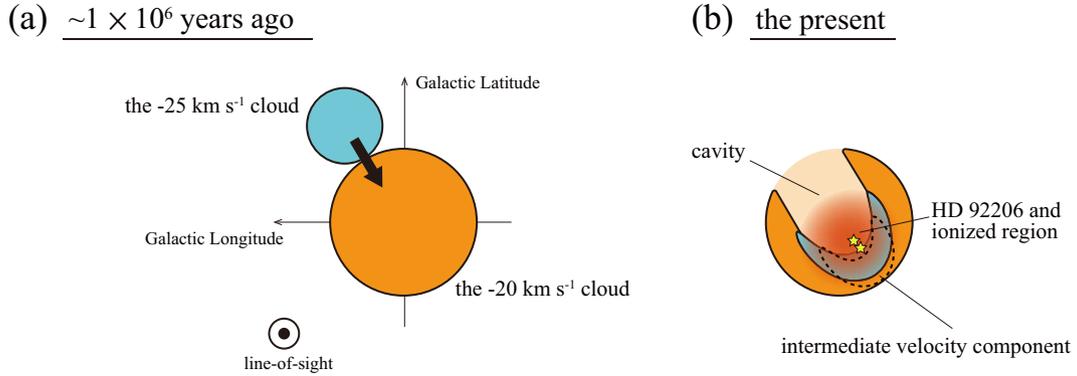}
  \end{center}
  \caption{A sketch diagram of the cloud--cloud collision scenario in Gum 31. }\label{fig:sche}
\end{figure*}

In Figure\,\ref{fig:lvs}(c), we found that a V-shape structure in the $l$--$v$ diagrams can be seen.
In addition, it was found that the two velocity components show complementary spatial distributions (Figure\,\ref{fig:red_blue}).
These morphology and velocity structures are very similar to the numerical simulations conducted by e.g., \citet{1992PASJ...44..203H}, \citet{2014ApJ...792...63T}, and \citet{2015MNRAS.450...10H}.
That is, a small cloud (the $-25$\,km\,s$^{-1}$ cloud) collided with the large cloud (the $-20$\,km\,s$^{-1}$ cloud) at supersonic velocity \textcolor{black}{from the eastern side,} and formed a cavity.
\textcolor{black}{The small cloud has a larger mass than the large cloud ($4.9\,\times \,10^4$\,$M_{\odot}$ and $2.3\,\times \,10^4$\,$M_{\odot}$, respectively). 
It is considered that the initial density of the small cloud was higher than that of the large cloud, if we assume that their thickness on the line-of-sight direction are about the same. 
These condition are similar to the numerical simulation conducted by \citet{2014ApJ...792...63T}.
After the collision, the two were merged in the interface layers}, and turbulence was predominant due to strong compression, forming a molecular cloud core that formed massive stars in HD 92206.
Furthermore, the collision scenario can interpret the offset of HD 92206 from the center of Gum 31 seen in Figure\,\ref{fig:RGB}(a) and the first moment \textcolor{black}{and spectra} of the molecular gas in Figures\,\ref{fig:mom1}.
Figure\,\ref{fig:sche} shows a sketch diagram of this collision scenario in Gum 31. 
Such a CCC scenario has also been proposed for RCW 120 (\cite{2015ApJ...806....7T}) and N4 (\cite{2019ApJ...872...49F}), which have a similar appearance to Gum 31.
The time scale of the collision is roughly estimated as $\sim$10 pc/(5\,km\,s$^{-1}\,\times \,$2)\,$\sim$\,1\,Myr from the size of the cavity and the velocity separation of the molecular cloud, with the assumption that the angle between the line of sight and the collision axis is $60^{\circ}$.
\textcolor{black}{There is no large contradiction between this collision time scale and the young age of \textcolor{black}{the cluster NGC 3324} ($< \, 3$\,Myr, \cite{2001A&A...371..107C}).}

\textcolor{black}{We conclude that the massive star formation of the CNC and Gum 31 were triggered by multiple CCCs and a CCC, respectively, because the CCC scenario in Gum31 is consistent with Paper I results in terms of the collision timescale and the collision direction. 
It is considered that the source that caused the multiple CCCs in the CNC and Gum31 is possibly the same. }
This picture is similar to the star formation history in W51 (\cite{2019PASJ..tmp...46F}), which has similar star formation activity \textcolor{black}{(dozens of O-type stars)} to Carina.
\textcolor{black}{In addition, our results support the proposal that Gum 31 is part of the CNC (\cite{2013A&A...552A..14O}).}

\textcolor{black}{For the YSOs (triangle markers in Figure\,\ref{fig:RGB}(a)), it cannot be determined by this study whether the formation was triggered by the collision or triggered by the feedback of stars in Gum 31 as suggested by \citet{2008A&A...477..173C}.
In the future, it will be possible to discuss this by e.g., detailed measurement of their ages.}

\section{Summary}
\begin{enumerate}
\item We analysed the $^{12}$CO ($J$=1--0), $^{13}$CO ($J$=1--0) (Mopra), and $^{12}$CO ($J$=2--1) (NANTEN2) emission lines toward Gum 31, and found the discrete three molecular clouds at velocities $-25$, $-20$, and $-10$ \,km\,s$^{-1}$.
\item The $-25$\,km\,s$^{-1}$ cloud and the $-20$\,km\,s$^{-1}$ cloud are associated with Gum 31, because their $^{12}$CO ($J$=2--1)/$^{12}$CO ($J$=1--0) intensity ratios are high. In addition, these clouds show the observational signatures of a CCC as same as the clouds in the CNC. In addition, their morphology and velocity structures are very similar to the numerical simulations conducted by the previous studies.
\item We propose a scenario that the $-25$\,km\,s$^{-1}$ cloud and the $-20$\,km\,s$^{-1}$ cloud were collided and triggered the formation of the massive stars in Gum 31. This \textcolor{black}{CCC} scenario can explain the offset of the stars from the center and the morphology of Gum 31 simultaneously. We estimate the timescale of the collision to be $\sim$\,1\,Myr by using the path length of the collision and the assumed velocity separation. This is \textcolor{black}{roughly} consistent with that of the CCCs in the CNC \textcolor{black}{estimated} in our previous study\textcolor{black}{, which support that Gum 31 is part of the CNC}. 
\end{enumerate}

\begin{ack}
We are deeply grateful to the financially support by Grants-in-Aid for Scientific Research (KAKENHI) of the Japanese society for the Promotion of Science (JSPS; grant numbers 15K17607,  17H06740, and 18K13580), and we would like to thank the all members of the NASA, Mopra, and NANTEN2 for providing the data. 
Also, we want to thanks Astropy (\cite{2013A&A...558A..33A}) and APLpy (\cite{2012ascl.soft08017R}) development team for data analysis. 
\end{ack}

\end{document}